\documentclass[aps,prl,reprint,10pt,twocolumn,superscriptaddress,floatfix]{revtex4-1}

\usepackage{times}
\usepackage{amsmath,amssymb,bm,graphicx}
\usepackage[breaklinks,colorlinks=true,urlcolor=blue,citecolor=blue,linkcolor=blue]{hyperref}
\usepackage{mathtools}
\usepackage{xcolor}
\usepackage{enumitem}

\DeclareMathOperator{\Pf}{Pf}

\begin{document}
\title{Numerical exploration of trial wave functions for the particle-hole-symmetric Pfaffian}

\date{\today}

\author{Ryan V. Mishmash}
\affiliation{Department of Physics and Institute for Quantum Information and Matter, California Institute of Technology, Pasadena, California 91125, USA}
\affiliation{Walter Burke Institute for Theoretical Physics, California Institute of Technology, Pasadena, California 91125, USA}
\affiliation{Department of Physics, Princeton University, Princeton, New Jersey 08540, USA}

\author{David F. Mross}
\affiliation{Department of Condensed Matter Physics, Weizmann Institute of Science, Rehovot, 76100, Israel}

\author{Jason Alicea}
\affiliation{Department of Physics and Institute for Quantum Information and Matter, California Institute of Technology, Pasadena, California 91125, USA}
\affiliation{Walter Burke Institute for Theoretical Physics, California Institute of Technology, Pasadena, California 91125, USA}

\author{Olexei I. Motrunich}
\affiliation{Department of Physics and Institute for Quantum Information and Matter, California Institute of Technology, Pasadena, California 91125, USA}
\affiliation{Walter Burke Institute for Theoretical Physics, California Institute of Technology, Pasadena, California 91125, USA}

\begin{abstract}
We numerically assess model wave functions for the recently proposed particle-hole-symmetric Pfaffian (`PH-Pfaffian') topological order, a phase consistent with the recently reported thermal Hall conductance [Banerjee et al., \href{https://www.nature.com/articles/s41586-018-0184-1}{Nature {\bf 559}, 205 (2018)}] at the ever enigmatic $\nu=5/2$ quantum-Hall plateau. We find that the most natural Moore-Read-inspired trial state for the PH-Pfaffian, when projected into the lowest Landau level, exhibits a remarkable numerical similarity on accessible system sizes with the corresponding (compressible) composite Fermi liquid. Consequently, this PH-Pfaffian trial state performs reasonably well energetically in the half-filled lowest Landau level, but is likely not a good starting point for understanding the $\nu=5/2$ ground state.  Our results suggest that the PH-Pfaffian model wave function either encodes anomalously weak $p$-wave pairing of composite fermions or fails to represent a gapped, incompressible phase altogether.
\end{abstract}

\maketitle

{\bf \emph{Introduction.}}~The half-filled Landau level has long stood as a paradigmatic example of an inherently strongly interacting quantum many-body system displaying various exotic phenomena. For the half-filled lowest Landau level (LLL), it is now experimentally \cite{willett_experimental_1993, kang_how_1993, goldman_detection_1994, smet_magnetic_1996} and numerically \cite{rezayi_fermi-liquid-like_1994, rezayi_incompressible_2000, geraedts_half-filled_2016} well established that the Coulomb-interacting ground state exhibits a Fermi sea of composite fermions \cite{jain_composite-fermion_1989, jain_composite_2007}---emergent degrees of freedom each consisting of an electron bound to two fictitious flux quanta that, on average, cancel the applied magnetic field. The theoretical description for this remarkable gapless `composite Fermi liquid' (CFL) phase was pioneered by Halperin, Lee, and Read (HLR) \cite{halperin_theory_1993}. Recently, the issue of particle-hole (PH) symmetry resurfaced as an important aspect of the problem following Son's proposal that composite fermions are Dirac particles \cite{son_is_2015}---a picture that has since found numerical support \cite{geraedts_half-filled_2016, balram_nature_2016, fremling_trial_2018, geraedts_berry_2017, yang_dirac_2017}.

An even subtler topic concerns the nature of the  $\nu=5/2$ plateau seen at the half-filled second Landau level (SLL) \cite{willett_observation_1987}.  Here, numerics generally support \cite{morf_transition_1998, rezayi_incompressible_2000, storni_fractional_2010} either Moore-Read (MR) Pfaffian topological order \cite{moore_nonabelions_1991} that emerges upon $p_x-ip_y$ pairing composite fermions \cite{read_paired_2000}, or its particle-hole conjugate, the anti-Pfaffian \cite{levin_particle-hole_2007, lee_particle-hole_2007}.  The Coulomb-interacting problem projected into the SLL exhibits an exact particle-hole symmetry; in this setting the MR-Pfaffian and anti-Pfaffian are exactly degenerate, and the emergence of one over the other requires spontaneous symmetry breaking \cite{peterson_spontaneous_2008, wang_particle-hole_2009}.  For more experimentally realistic models that include, for example, Landau level mixing (which explicitly breaks particle-hole symmetry), it is now believed---after some debate \cite{wojs_landau-level_2010, pakrouski_phase_2015, rezayi_breaking_2011, zaletel_infinite_2015}---that the anti-Pfaffian state is favored microscopically \cite{rezayi_landau_2017}.  

The experimental status of the $\nu=5/2$ quantum Hall state is similarly complex; for a review, see Ref.~\cite{willett_quantum_2013}. In a recent breakthrough, Banerjee et al.~\cite{banerjee_observation_2018} measured the thermal Hall conductance and found $\kappa_{xy}\approx 5/2$ [in units of $\frac{\pi^2 k_B^2}{3h}T$]. Assuming all edge modes have equilibrated, $\kappa_{xy}$ probes the edge's total chiral central charge \cite{kane_quantized_1997}; a half-integer value directly implies an odd number of Majorana edge modes and concomitant non-Abelian Ising anyon bulk quasiparticles \cite{nayak_non-abelian_2008}. Intriguingly, $\kappa_{xy}=5/2$ is half-integer yet corresponds to the edge structure of neither the MR-Pfaffian nor the anti-Pfaffian, but rather is consistent with \emph{particle-hole-symmetric Pfaffian} (PH-Pfaffian) topological order
recently elucidated by Son \cite{son_is_2015}.  (Importantly, PH-Pfaffian topological order is \emph{compatible} with particle-hole symmetry, but does not require it \cite{zucker_stabilization_2016}; cf.~Refs.~\cite{chen_symmetry_2014, bonderson_time-reversal_2013}.)  This result is confounding, particularly in light of the extensive numerical work summarized above, which has not revealed any evidence for such a state.  The observed $\kappa_{xy}=5/2$ can nevertheless be plausibly explained by disorder-induced PH-Pfaffian behavior \cite{mross_theory_2017, wang_topological_2017, lian_theory_2018} or incomplete thermal equilibration of an anti-Pfaffian edge \cite{simon_interpretation_2018}.

In this paper we focus on the clean limit and numerically explore minimal trial PH-Pfaffian wave functions projected into a fixed Landau level.  We specifically extract their degree of particle-hole symmetry, pair correlation functions, entanglement spectra, and overlap with exact ground states of model Hamiltonians and other trial wave functions, in systems with up to $N = 12$ electrons.  These diagnostics all reveal a striking similarity between our trial PH-Pfaffian wave functions and, surprisingly, the compressible CFL state.  We discuss several possible interpretations of these results and associated conundrums that they raise.

{\bf \emph{Trial wave functions.}}~The hallmark of PH-Pfaffian topological order is a reversed chirality of the Majorana edge mode relative to that of the MR-Pfaffian. Therefore, in analogy with the celebrated MR-Pfaffian wave function \cite{moore_nonabelions_1991},
\begin{equation}
\Psi_\mathrm{MR-Pf}(\{z_i\}) = \Pf\left[\frac{1}{z_i - z_j}\right]\prod_{i<j}(z_i - z_j)^2,
\label{eq:MR-Pf}
\end{equation}
we may view the PH-Pfaffian as a $p_x+ip_y$ superconductor of composite fermions \cite{son_is_2015} naturally described by \cite{zucker_stabilization_2016,yang_particle-hole_2017}
\begin{equation}
\Psi_\mathrm{PH-Pf}(\{z_i\}) = \mathcal{P}_\mathrm{LLL}\Pf\left[\frac{1}{z_i^* - z_j^*}\right]\prod_{i<j}(z_i - z_j)^2.
\label{eq:PH-Pf}
\end{equation}
[We always work in the LLL Hilbert space, and for brevity, we omit the Gaussian factors $e^{-\frac{1}{4}\sum_i |z_i|^2}$. SLL wave functions (pertinent to $\nu=5/2$) can formally be obtained by applying raising operators to the LLL wave functions.]
Importantly, the presence of antiholomorphic terms on the right-hand side of Eq.~\eqref{eq:PH-Pf} necessitates an explicit LLL projection, contrary to the MR-Pfaffian.
Our primary goal is to numerically characterize the trial PH-Pfaffian wave function in Eq.~\eqref{eq:PH-Pf}, assessing its efficacy for describing gapped PH-Pfaffian topological order in the half-filled Landau level.  To this end it will prove very useful to also consider a model CFL wave function based on the HLR construction \cite{rezayi_fermi-liquid-like_1994} (also putatively able to capture some aspects of Dirac composite fermions \cite{balram_nature_2016, fremling_trial_2018, geraedts_berry_2017, wangsenthil2016, ChongPHS, kumar_composite_2018}):
\begin{equation}
\Psi_\mathrm{CFL}(\{z_i\}) = \mathcal{P}_\mathrm{LLL}\det[e^{i \mathbf{k}_j\cdot\mathbf{r}_i}]\prod_{i<j}(z_i - z_j)^2,
\label{eq:CFL}
\end{equation}
with $\det[e^{i \mathbf{k}_j\cdot\mathbf{r}_i}]$ a Slater determinant of plane-wave orbitals.

For our numerics, we consider a spherical geometry \cite{haldane_fractional_1983} in which $N$ electrons moving on the surface experience a radial magnetic field produced by a monopole of strength $Q>0$ at the origin. The sphere has radius $\sqrt{Q}\,\ell_0$, with $\ell_0=\sqrt{\hbar c/e B}$ the magnetic length, and $N_\phi=2Q$ flux quanta penetrate its surface. Quantum-Hall states defined in the plane, such as Eqs.~\eqref{eq:MR-Pf} through \eqref{eq:CFL}, can be translated to the sphere using a standard procedure \cite{haldane_fractional_1983, fano_configuration-interaction_1986}.  Due to finite curvature of the sphere, states at a given filling  $\nu$ are characterized by their `shift' quantum number $\mathcal{S}$ \cite{wen_shift_1992} via $N_\phi = \nu^{-1}N - \mathcal{S} = 2N - \mathcal{S}$ (taking $\nu=1/2$). The MR-Pfaffian and PH-Pfaffian states respectively occur at shifts $\mathcal{S}_\mathrm{MR} = 3$ and $\mathcal{S}_\mathrm{PH} = 1$.  Note that the latter shift corresponds to a LLL Hilbert space of dimension $N_\mathrm{orb} = N_\phi + 1 = 2N$, a clear zeroth-order condition for the possibility of particle-hole symmetry.

We can define CFL states on the sphere following Rezayi and Read \cite{rezayi_fermi-liquid-like_1994}. Specifically, upon attaching two flux quanta to each electron that oppose the external field, each such composite fermion feels a total average magnetic flux of $2q = N_\phi - 2(N-1)$. Thus, at the respective shifts $\mathcal{S}_{\rm MR}$ and $\mathcal{S}_{\rm PH}$, we have $q_\mathrm{MR}=-1/2$ and $q_\mathrm{PH}=+1/2$ \footnote{At the shift $\mathcal{S}=2$, composite fermions feel identically zero magnetic field at $\nu=1/2$ so that $q=0$ \cite{rezayi_fermi-liquid-like_1994}.}. CFL wave functions are then obtained by replacing the plane waves in Eq.~\eqref{eq:CFL} by appropriate monopole harmonics \cite{wu_dirac_1976, wu_properties_1977}: $\det[e^{i \mathbf{k}_j\cdot\mathbf{r}_i}] \to \det[Y^q_{\ell_j,m_j}(\theta_i, \phi_i)]$. 

\begin{figure}[t]
  \begin{center}
    \includegraphics[width=0.9\columnwidth]{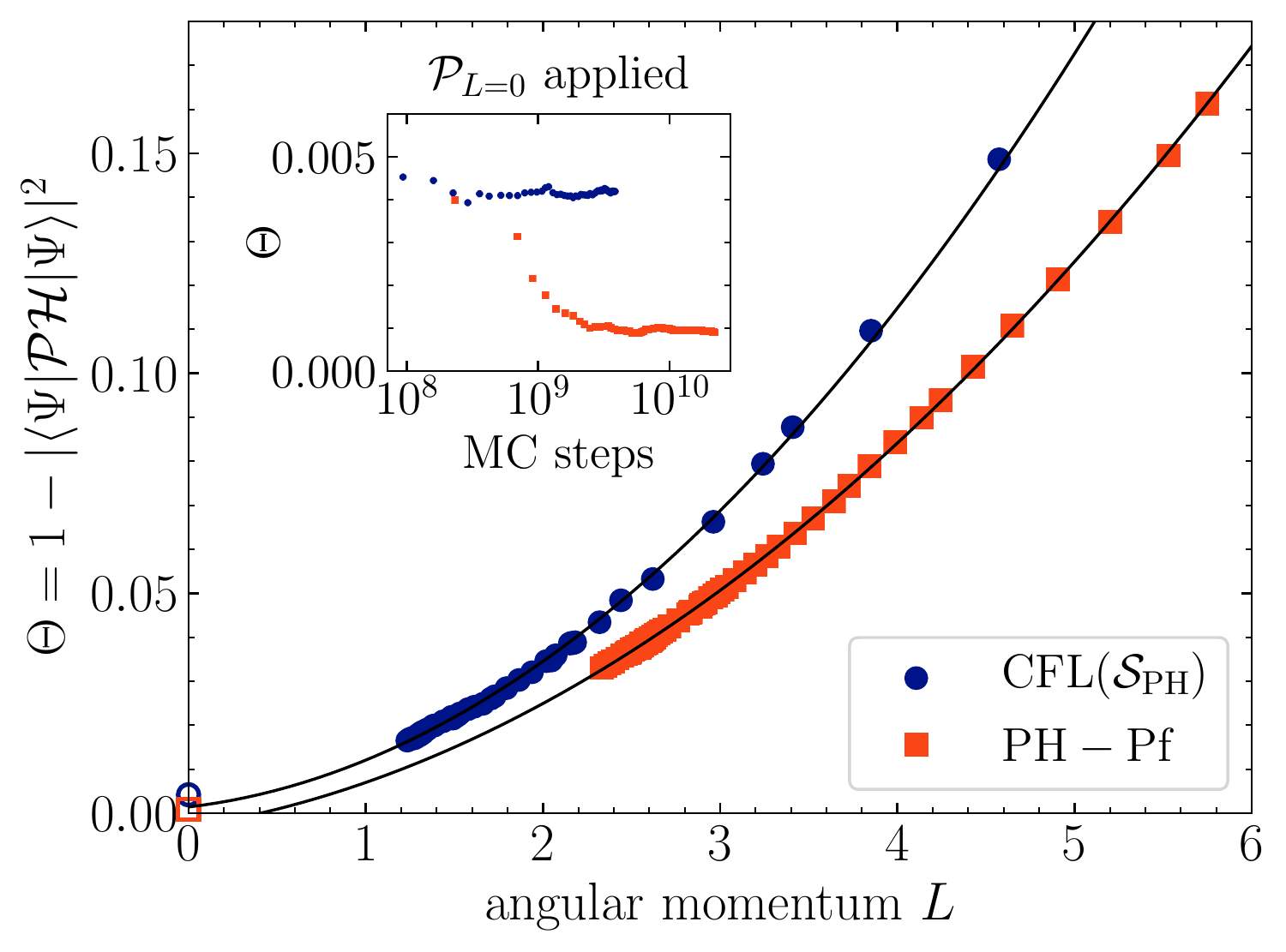}
  \end{center}
  \vspace{-0.2in}
  \caption{Measured particle-hole asymmetry $\Theta$ versus measured total angular momentum $L$ for MC evaluation of the $N=12$ CFL($\mathcal{S}_{\rm PH}$) and PH-Pfaffian trial states. Different solid points correspond to different numbers of MC samples, while the open points at $L=0$ represent the best MC state projected to $L=0$; the solid curves are fits to a quadratic polynomial. In the inset, we show $\Theta$ \emph{after projecting to $L=0$} versus total number of MC steps.}
  \label{fig:PHanal}
\end{figure}

\begin{figure}[t]
  \begin{center}
    \includegraphics[width=1.0\columnwidth]{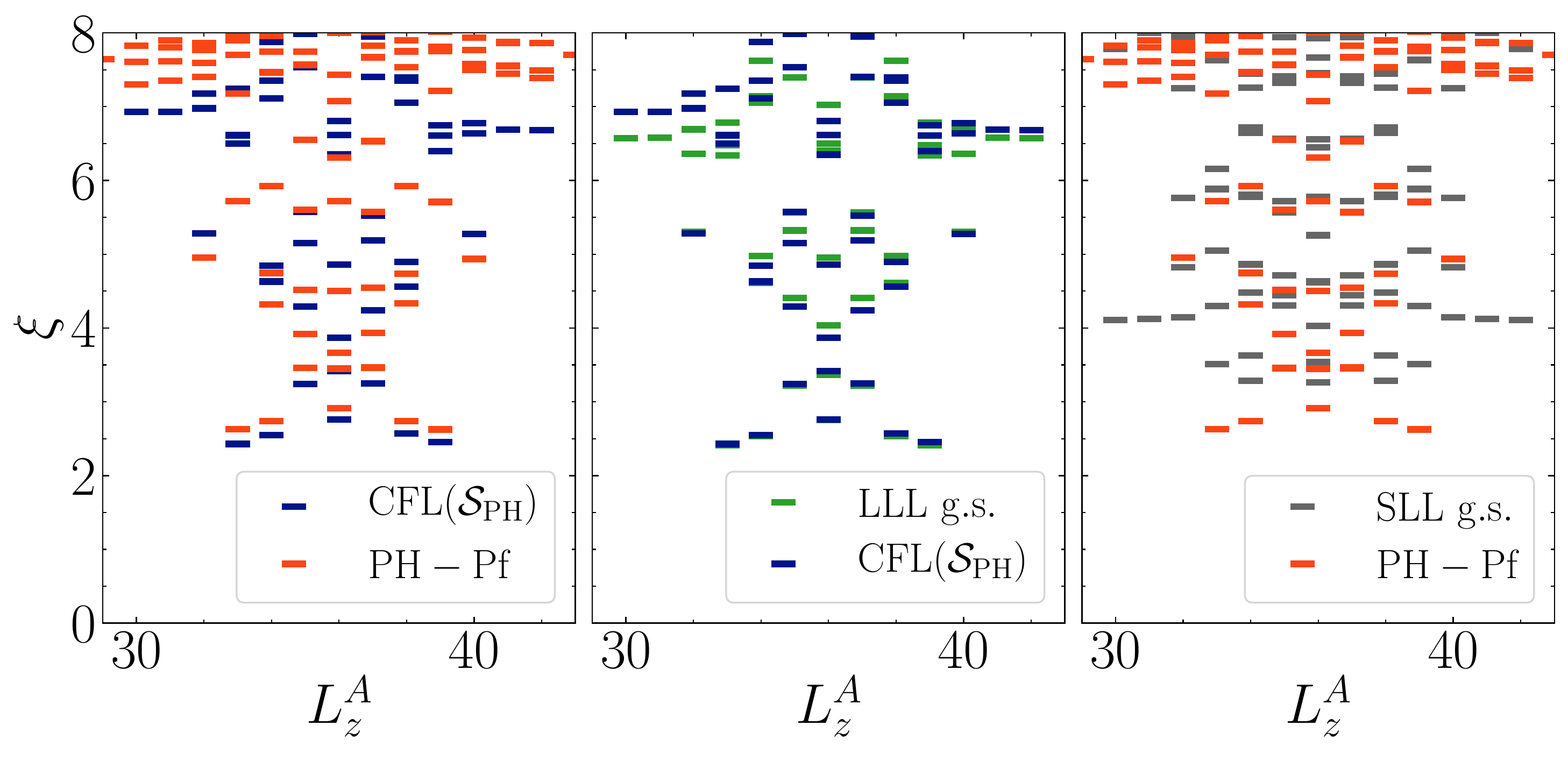}
  \end{center}
  \vspace{-0.2in}
  \caption{Orbital-partition entanglement spectra for the $N=12$, $N_\mathrm{orb} = N_\phi + 1 =24$ system with $N_\mathrm{orb}^A = 12$ and $N^A = 6$. In the left panel, we depict the close similarity between the CFL($\mathcal{S}_{\rm PH}$) and PH-Pfaffian trial states, while in the other two panels we compare the respective trial states to the LLL (middle) and SLL (right) Coulomb ground states.}
  \label{fig:ent-spec}
\end{figure}

We calculate via Monte Carlo (MC) sampling all coefficients of the trial wave functions in the many-body Fock space built out of LLL orbitals on the sphere, i.e., $\langle\{m_i\}|\Psi_\mathrm{trial}\rangle$ where $m_i$ is the $z$-component of angular momentum of particle $i$.  There are two reasons for employing this numerically challenging, brute-force approach (see also Ref.~\cite{balram_nature_2016}).  First, it is a completely general way to perform LLL projection; notably, Eq.~\eqref{eq:PH-Pf} is not amenable to the Jain-Kamilla projection scheme \cite{jain_quantitative_1997} thereby severely limiting the accessible system sizes.
Second, obtaining the full second-quantized wave function constitutes the only practical way to calculate important quantities such as the degree of particle-hole symmetry (Fig.~\ref{fig:PHanal}) and the entanglement spectrum (Fig.~\ref{fig:ent-spec}).

In what follows, we closely compare PH-Pfaffian and CFL model wave functions at shift $\mathcal{S}_{\rm PH}$, i.e., with $N_\phi = 2N-1$.  Hereafter we will denote the CFL at shift $\mathcal{S}$ by CFL($\mathcal{S}$). The CFL($\mathcal{S}_{\rm PH}$) state has filled angular momentum shells at electron numbers $N=2,6,12,20,\dots$ \footnote{Filled shells also occur at these $N$ for CFL($\mathcal{S}_{\rm MR}$) since $|q_\mathrm{MR}| = |q_\mathrm{PH}| = 1/2$.}. The largest such system amenable to our MC approach has $N=12$ and $N_\phi=23$---hence, we focus extensively on this system. In this case, the many-body Hilbert space in the $L_z = \sum_i m_i = 0$ sector contains 61108 basis states. Fully diagonalizing the total-angular-momentum operator $\hat{\mathbf{L}}^2$ in this basis reveals that there are $\dim\mathcal{H}_{L=0} = 127$ eigenstates with $L=0$. We use these states to form a projection operator $\mathcal{P}_{L=0}$ that we apply to our MC-acquired wave functions to obtain final, perfectly rotationally invariant trial states.  While the above trial wave functions adapted to the sphere are exact $\hat{\mathbf{L}}^2$ eigenstates with $L = 0$ \cite{fano_configuration-interaction_1986}, statistical error introduced by our MC scheme spoils this property; $\mathcal{P}_{L=0}$ merely removes this error.  

{\bf \emph{Physical properties.}}~Although the CFL($\mathcal{S}_{\rm PH}$) and PH-Pfaffian trial wave functions exist at the `correct' shift, their degree of particle-hole symmetry on finite-size systems is not manifest. Figure~\ref{fig:PHanal} illustrates the particle-hole asymmetry $\Theta = 1-|\langle\Psi|\mathcal{PH}|\Psi\rangle|^2$ exhibited by these wave functions as obtained via MC for the $N=12$ system ($\mathcal{PH}$ denotes particle-hole conjugation with respect to the fully filled Landau level). The main panel plots measured $\Theta$ versus measured $L$ [defined through $\langle\Psi|\hat{\mathbf{L}}^2|\Psi\rangle=L(L+1)$] for different MC runs with varying numbers of MC steps \emph{before applying $\mathcal{P}_{L=0}$}; solid curves depict fits to a second-order polynomial. In the inset, we monitor $\Theta$ versus MC steps \emph{after applying $\mathcal{P}_{L=0}$}. These results indicate $\Theta = O(10^{-3})$ for both wave functions.  At smaller sizes $N=4,6,8,10$ we find for the PH-Pfaffian trial state $\Theta=0, 3.3\times10^{-6}, 2.9\times10^{-4}, 2.1(2)\times10^{-4}$ (exact evaluation is possible for $N\leq8$). The degree of particle-hole symmetry shown by both wave functions up to $N=12$ is impressively high; see Refs.~\cite{yang_particle-hole_2017, geraedts_berry_2017, yang_dirac_2017} for recent related discussions. 
Still, we expect particle-hole symmetry to be completely lost in the thermodynamic limit since the subset of particle-hole-symmetric wave functions has negligible measure within the set of all possible wave functions in the same phase. Given that $\Psi_\mathrm{PH-Pf}$ lacks exact particle-hole symmetry for $N>4$ and in this sense describes a `generic' wave function, we conclude that $\Theta\to1$ is the most likely scenario. (This is also consistent with the monotonous growth of $\Theta$ with $N$ at small $N$.)  

\begin{figure}[t]
  \begin{center}
    \includegraphics[width=0.9\columnwidth]{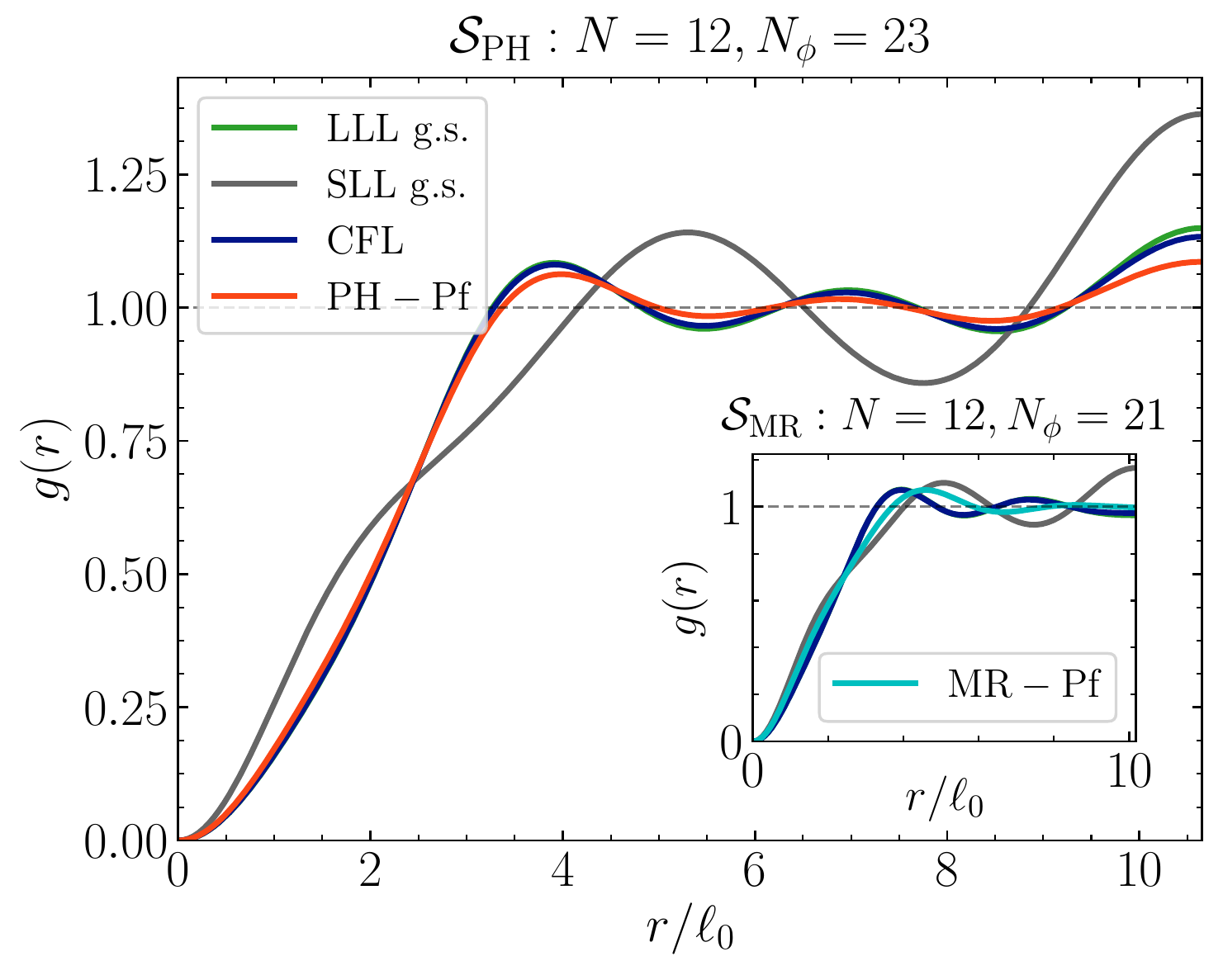}
  \end{center}
  \vspace{-0.2in}
  \caption{Pair correlation function $g(r)$ for the LLL and SLL Coulomb ground states as well as the CFL($\mathcal{S}_{\rm PH}$) and PH-Pfaffian trial states at the shift $\mathcal{S}_{\rm PH}$ (main panel). The inset shows data for the analogous states at $\mathcal{S}_{\rm MR}$; the first three legend entries in the main panel also apply to the inset.}
  \label{fig:gr}
\end{figure}

Next, we examine the `pair correlation function' evaluated along the sphere's equator: $g(r) = \frac{1}{\rho^2}\langle \hat{\psi}^\dagger(r) \hat{\psi}^\dagger(0) \hat{\psi}(0) \hat{\psi}(r) \rangle$, where $\rho$ is the 2D density and $\hat{\psi}$ is the LLL-projected electron operator.  The main panel of Fig.~\ref{fig:gr} corresponds to shift $\mathcal{S}_{\rm PH}$.  Specifically, we show data for PH-Pfaffian and CFL($\mathcal{S}_{\rm PH})$ trial states, and for the ground state of the Coulomb potential (defined in terms of the chord distance) projected into either the LLL or the SLL (implemented via Haldane pseudopotentials \cite{haldane_fractional_1983, toke_understanding_2006, wooten_haldane_2013}). We again focus on $N=12$ and apply $\mathcal{P}_{L = 0}$ to all MC-obtained trial states \footnote{The norm of the wave functions post projection to $L=0$ is greater than 0.99 (for subsequent measurements we renormalize). Upon applying $\mathcal{P}_{L=0}$, the MC error for all data that we present is on the order of the symbol size or smaller, whereas without projection this would not be the case.}. Remarkably, the PH-Pfaffian and CFL($\mathcal{S}_{\rm PH})$ data are qualitatively indistinguishable. In fact, these two trial states exhibit very high overlap: $|\langle\Psi_{\mathrm{CFL}}|\Psi_\mathrm{PH-Pf}\rangle|^2=0.9106(2)$. At long distances, we expect $g(r)$ to approach unity as an oscillatory exponential (power law) for a gapped (gapless) phase. While the asymptotic behavior exhibited by the PH-Pfaffian trial state is not obvious at these sizes (it does have slightly reduced oscillation amplitudes at the largest distances), its similarity with the CFL wave function calls into question whether $\Psi_{\rm PH-Pf}$ represents a gapped phase \cite{milovanovic_paired_2017}. Finally, the data for the LLL ground state unsurprisingly closely tracks the PH-Pfaffian and especially CFL trial states \footnote{At the shift $\mathcal{S}_{\rm PH}$, we have checked (for up to $N=15$ electrons) that the total angular momentum of the LLL ground state follows the `shell filling' prediction (Hund's second rule) \cite{rezayi_fermi-liquid-like_1994} for composite fermions in the presence of a $q_\mathrm{PH}=+1/2$ monopole.}, while that for the SLL ground state behaves very differently.

For comparison, the inset of Fig.~\ref{fig:gr} presents analogous $g(r)$ data taken at shift $\mathcal{S}_{\rm MR}$ with $N=12$ and $N_\phi=21$. The MR-Pfaffian and CFL($\mathcal{S}_{\rm MR})$ data differ substantially as expected, and the overlap of the two wave functions is significantly reduced to $0.384(4)$ even though the $L=0$ Hilbert space contains fewer states than at shift $\mathcal{S}_{\rm PH}$ ($\dim\mathcal{H}_{L=0} = 52$ versus 127). Additionally, $g(r)$ for the MR-Pfaffian is trending in the direction of the SLL ground state while saturating to $g(r)\to1$ at long distances. Note that while the properties of the \emph{trial states} are reasonably well converged already at $N=12$, clearly observing incompressibility in the Coulomb-interacting SLL system itself at $\mathcal{S}_\mathrm{MR}$ requires $N\geq20$ electrons \cite{feiguin_density_2008}.

First introduced by Ref.~\cite{li_entanglement_2008} in the MR-Pfaffian context, the entanglement spectrum (ES) has since become an invaluable tool for characterizing topological phases. Figure~\ref{fig:ent-spec} presents orbital-partition entanglement spectra for the same wave functions in the main panel of Fig.~\ref{fig:gr}.  We work at shift $\mathcal{S}_{\rm PH}$ with $N=12$ electrons and $N_\mathrm{orb} = N_\phi + 1 = 24$ total orbitals; subsystem $A$ is chosen as the 12 states with positive angular momentum in the $z$-direction (which have predominant weight in the upper hemisphere). In Fig.~\ref{fig:ent-spec} we show the sector corresponding to $N^A = N/2 = 6$ electrons in subsystem $A$ and plot the `entanglement energies' $\xi$ versus $L_z^A$, the total subsystem $z$-component angular momentum \footnote{We use the same conventions for $\xi$ as Ref.~\cite{li_entanglement_2008}.}. A particle-hole transformation on a given wave function takes $L_z^A \to L_z^\mathrm{max} - L_z^A$, where $L_z^\mathrm{max} = \frac{1}{2}\left(\frac{1}{2} + Q\right)^2$ is the maximum total $z$-component angular momentum of the entire system. For the data in Fig.~\ref{fig:ent-spec} with monopole strength $Q=N_\phi/2=23/2$, we have $L_z^\mathrm{max} = 72$; hence, perfect particle-hole symmetry implies a reflection symmetry in the ES data about $L_z^A = L_z^\mathrm{max}/2 = 36$.

Interpreting the ES of PH-Pfaffian topological order poses an interesting open question.  Here we simply observe that for PH-Pfaffian model wave functions with up to $N = 12$ electrons, the ES exhibits a high degree of symmetry about $L_z^\mathrm{max}/2$ and closely tracks the CFL($S_{\rm PH}$) ES at the lowest $\xi$.  These properties are subtler manifestations of the near particle-hole symmetry of $\Psi_{\rm PH-Pf}$ and its high overlap with $\Psi_{\rm CFL}$ captured earlier.  Figure~\ref{fig:ent-spec} also shows comparisons between the CFL($\mathcal{S}_{\rm PH}$) trial state and LLL Coulomb ground state [which have near unit overlap: 0.9926(1)] and between the PH-Pfaffian trial state and SLL Coulomb ground state [which have near zero overlap: 0.01811(6)].  The pure Coulomb interaction projected into the SLL likely sits at a first-order phase transition between the MR-Pfaffian and anti-Pfaffian \cite{wang_particle-hole_2009}, which would naturally explain the lack of clear structure in the ES for that case.

\begin{figure}[t]
  \begin{center}
    \includegraphics[width=0.9\columnwidth]{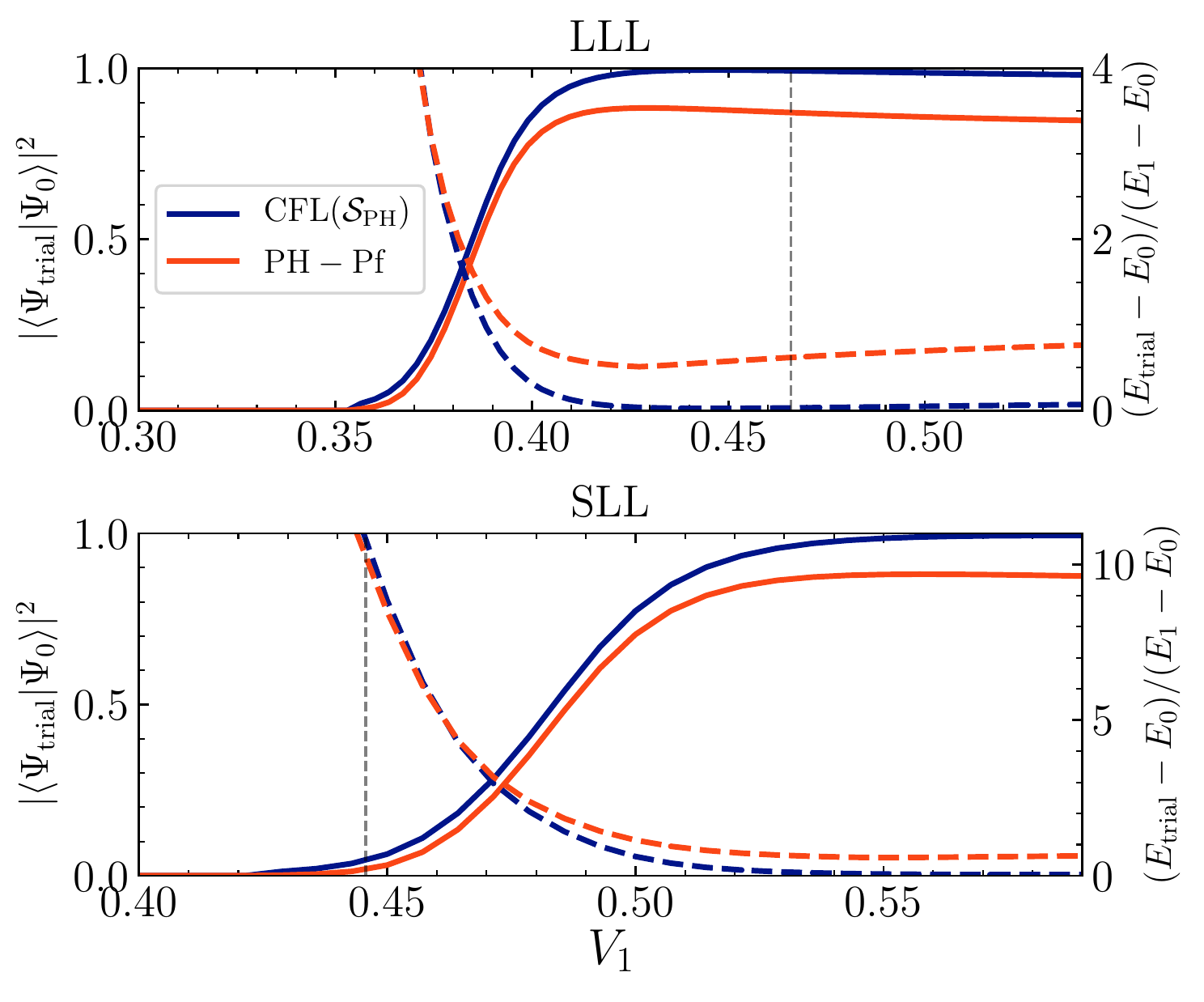}
  \end{center}
  \vspace{-0.2in}
  \caption{Overlaps (left axes, solid curves) and energies (right axes, dashed curves) of the CFL($\mathcal{S}_{\rm PH}$) and PH-Pfaffian trial states compared with the exact ground state $|\Psi_0\rangle$ at the shift $\mathcal{S}_{\rm PH}$ for Coulomb pseudopotentials in the LLL (top panel) and SLL (bottom panel) but with variable $V_1$. As in Figs.~\ref{fig:PHanal} through \ref{fig:ent-spec}, here $N=12$; the vertical dashed lines indicate the respective pure Coulomb points.}
  \label{fig:energies_overlaps}
\end{figure}

The similarity between the PH-Pfaffian and CFL$(\mathcal{S}_{\rm PH})$ trial states uncovered above suggests that the LLL-projected Coulomb interaction may be a reasonable starting point for realizing the former model wave function as the best approximation of the ground state compared to other natural trial states.  As a preliminary exploration, Fig.~\ref{fig:energies_overlaps} shows variational energies (relative to the ground-state energy, $E_0$, and normalized by the gap, $E_1-E_0$) and overlaps with the exact ground state for both CFL$(\mathcal{S}_{\rm PH})$ and PH-Pfaffian trial states.  Here we vary the pseudopotential $V_1$ and keep all other pseudopotentials fixed at their Coulomb values in the LLL (top panel) and SLL (bottom panel).  In the LLL, adding short-distance attraction to the potential by decreasing $V_1$ relative to its Coulomb value slightly improves the PH-Pfaffian trial energy and overlap, but fails to overcome CFL$(\mathcal{S}_{\rm PH})$ in this parameter regime.  We reach similar conclusions upon varying $V_3$ in addition to $V_1$. Finally, both wave functions perform extremely poorly near the Coulomb point for the SLL as relevant to the experimental $\nu=5/2$ plateau. That the PH-Pfaffian model wave function does poorly here is perhaps not surprising in light of the $N$-dependent ground state angular momentum of the pure 2D Coulomb SLL ground state at shift $\mathcal{S}_\mathrm{PH}$ \cite{wojs_transition_2009, luo_pfaffian_2017, balram_parton_2018}; therefore, we speculate that more drastic changes to the model Hamiltonian may be necessary.

{\bf \emph{Discussion.}}~A far more pressing matter concerns the nature of the PH-Pfaffian model wave function itself given similarities to the CFL.  The stark difference between MR-Pfaffian and CFL trial states at the same system sizes suggests that these similarities are not merely finite-size artifacts. We have also considered generalizations of $\Psi_{\rm PH-Pf}$ by including `stabilization' factors $\prod_{i<j}|z_i-z_j|^\alpha$ in Eq.~\eqref{eq:PH-Pf} before projection. Topological orders described by a $K$-matrix $K=\begin{pmatrix}n & m \\ m & n\end{pmatrix}$ with $n<m$ actually necessitate such factors for thermodynamic stability \cite{McDonaldHaldane1996, de_gail_plasma_2008}. For $\nu=5/2$, the 113 state is a plausible candidate that requires stabilization \cite{Feldman113}; a possible LLL stabilized wave function is 
\begin{align}
 \Psi_{113}(\{v_i,w_i\}) & = \mathcal{P}_\text{LLL} \prod_{i<j} |v_i-v_j|^4|w_i-w_j|^4 
 \nonumber\\
 \times & \prod_{i<j} (v_i-v_j)(w_i-w_j) \prod_{i,j} (v_i-w_j)^3
\end{align}
with $v_i,w_i$ complex coordinates for two species of distinguishable particles and $i \in [1,N/2]$. Reference~\cite{CDL} argued that the 113 and PH-Pfaffian topological orders share an intimate relation analogous to that between the Halperin 331 state \cite{Halperin1983} and the MR-Pfaffian \cite{moore_nonabelions_1991, read_paired_2000}. This relationship is encoded in the wave functions studied here: Fully antisymmetrizing the 331 wave function over all coordinates yields  $\Psi_\mathrm{MR-Pf}$ \cite{Greiter331Pfaffian}; similarly, associating $\{z_i\} = \{v_i\}$ and $\{z_{i+N/2}\} = \{w_i\}$, antisymmetrizing over all $N$ coordinates $z_{i=1,\ldots,N}$, and then LLL projecting the \textit{stabilized} wave function $\Psi_{113}$ yields $\Psi_{\rm PH-Pf}$ modified by $\prod_{i<j}|z_i-z_j|^\alpha$ with $\alpha=2$ \footnote{The modified PH-Pfaffian trial state with $\alpha=2$ coincides with the $k=2$ wave function constructed in Ref.~\cite{jolicoeur_non-abelian_2007}.}. Surprisingly, such factors very weakly affect the resulting LLL-projected PH-Pfaffian trial state; e.g., for $N=12$ electrons, wave functions with $\alpha=2$ and $\alpha=0$ [Eq.~\eqref{eq:PH-Pf}] have an overlap of 0.9931(2) \footnote{Reference~\cite{balram_nature_2016} observed that such factors weakly affect LLL composite fermion wave functions more generally.}. This broader family of states thus also appears closely related to the CFL.  

One logical possibility is that our PH-Pfaffian trial states describe gapped PH-Pfaffian topological order in the thermodynamic limit, but with pairing that is significantly suppressed by LLL projection.  This interpretation raises an interesting puzzle: What determines the anomalous pairing strength given the absence of any obvious small parameter in the PH-Pfaffian model wave functions? Another possibility is that LLL projection obliterates the pairing entirely, and that in the thermodynamic limit the PH-Pfaffian trial wave functions describe a gapless state in the same university class as the CFL.  Here, too, a conundrum arises.  Upon removing $\mathcal{P}_{\rm LLL}$, Eq.~\eqref{eq:PH-Pf} certainly describes gapped PH-Pfaffian topological order (without particle-hole symmetry).  In this scenario LLL projection would qualitatively alter the universal properties of the trial state---contrary to the typical situation---for possibly fundamental reasons that are presently unclear.  Landau-level mixing might then, counterintuitively, be \emph{required} to stabilize PH-Pfaffian topological order.  At present we cannot rule out the possibility that alternative LLL-projected PH-Pfaffian trial states do not suffer from the subtleties encountered here.  In this regard, it would be interesting to construct trial states for the PH-Pfaffian with additional variational freedom---one natural route is to consider more general pair wave functions for a $p_x+ip_y$ superconductor of composite fermions in the spirit of Ref.~\cite{moller_paired_2008}. 

\acknowledgments
\emph{Note added}: After completion of this work, Ref.~\cite{balram_parton_2018} appeared which contains some overlap with our results; see their Appendix A.

R.V.M.~gratefully acknowledges Mike Zaletel for valuable discussions; we thank Ajit Balram for pointing out the connection between the $\alpha=2$ generalized PH-Pfaffian state and the states proposed in Ref.~\cite{jolicoeur_non-abelian_2007}. This work was supported by grant No.~2016258 from the United States-Israel Binational Science Foundation (BSF); the Minerva foundation with funding from the Federal German Ministry for Education and Research (D.F.M.); the Army Research Office under Grant Award W911NF-17-1-0323 (J.A.); the NSF through grants DMR-1723367 (J.A.) and DMR-1619696 (O.I.M.); the Caltech Institute for Quantum Information and Matter, an NSF Physics Frontiers Center with support of the Gordon and Betty Moore Foundation through Grant GBMF1250; and the Walter Burke Institute for Theoretical Physics at Caltech.

\bibliography{PH-Pf}

\end{document}